# Measurement of the variable surface charge concentration in gallium nitride and implications on device modeling and physics

E. Ber, B. Osman, and D. Ritter

*Abstract*— We have evaluated the density of interface trap states ($D_{it}$) at the surface of a GaN/AlGaN/GaN heterojunction by the previously described gated van der Pauw experiments, as well as by a UV assisted gated van der Pauw method, described in this article. The obtained $D_{it}$ values are about two orders of magnitude lower than assumed by the surface-donor theory and three orders of magnitude lower than required to compensate the polarization surface charge in GaN. Previous experimental studies using a variety of other techniques reported similarly low $D_{it}$ values. We hence conclude that variable midgap surface-charge is not responsible for the formation of the two-dimensional electron gas, and cannot compensate for the large surface polarization charge in GaN. A yet unexplained polarization self-compensating (PSC) surface charge must be invoked to account for experiments. A few comments about the physical nature of the proposed PSC charge are provided.

*Index Terms*—Gallium nitride, GaN/AlGaN/GaN heterojunction, density of interface states, surface charge, polarization charge.

## I. Introduction

GALLIUM nitride surface charge plays a significant role in the device physics of heterostructure field effect transistors (HFETs). Standard modeling assumes three types of surface charge: fixed polarization charge, variable midgap surface charge, and an additional fixed surface charge.

The value of the polarization charge is not well established. It is difficult to measure it directly, yet light emission from stacking faults provided a polarization charge value of -0.022 C/m$^2$ [1], in agreement with earlier theory. A recent theory, however, finds that the polarization charge in GaN is 1.312 C/m$^2$ [2], namely about 60 times larger, and with an opposite polarity.

Variable surface charge is introduced into device modeling as a surface density of states, which can be either donor-like or acceptor-like. The occupation of surface states depends upon the position of the Fermi level at the surface. Application of an electric field or radiation changes the occupation of the surface states. The variable midgap surface charge is therefore accessible experimentally, and its concentration can be evaluated.

E. Ber, B. Osman, and D. Ritter are with the Department of Electrical Engineering, Technion-Israel Institute of Technology, Haifa 32000, Israel (e-mail: semanbar@campus.technion.ac.il)

An additional fixed surface charge is often required to model HFET devices. A fixed surface charge has no obvious physical origin, but is frequently introduced to obtain agreement with experiments. It was previously suggested that a deep border trap inside an insulator deposited on top of GaN is the origin of the fixed interface charge [3]. This model, however, does not apply for bare GaN surfaces.

The variation of the two-dimensional electron gas (2DEG) concentration in GaN heterojunctions with barrier thickness is well documented experimentally [4], [5]. It is explained by the surface donor model, which assumes a high concentration of donor-like surface-states at a discrete level, or across some energy band. When the barrier is sufficiently thick, the Fermi level resides within the surface-donor energy band, and a fraction of the donor-states is unoccupied, compensating the 2DEG negative charge. The surface donor model requires that the surface-donor concentration is larger than the 2DEG concentration [4], [6], [7], [8]. Otherwise, the unoccupied surface–donor charge cannot compensate the 2DEG charge.

The density of surface states in GaN was previously evaluated by a variety of methods, such as capacitance-voltage (CV) measurements, UV assisted CV measurements, capacitance-voltage curves measured at different temperatures, photo-assisted capacitance transient, and comparison of metal insulator semiconductor (MIS) transistor characteristics to simulations [3], [6], [9]–[16]. The outcome of most experiments was that the concentration of the variable surface-charge (i.e., the density of surface-states) is lower than the 2DEG concentration and much lower than the polarization charge.

To augment the existing experimental data, we report here on gated van der Pauw measurements of surface states in GaN [17], [18]. Previously, we have reported results only across a limited energy range. Here, we extend the results across the entire bandgap. To validate our data, we have introduced a new method; the UV assisted gated van der Pauw experiment. Using both methods, we found that the concentration of surface states is lower by about two orders of magnitude than required to model the variation of the 2DEG concentration with barrier thickness, and to explain how the polarization charge in nitrides is compensated. An additional surface charge must hence be identified to account for the above two phenomena. Some comments about the missing charge are presented in the discussion.

## II. Experimental

We first report on results obtained by the gated van der Pauw technique [17], [18]. Briefly, a voltage step is applied to an MIS gate while the 2DEG conductivity is monitored as a function of time. The time variation of the 2DEG conductivity is due to electron trapping (or de-trapping) from energy-states in the barrier and interface [19]. Two assumptions underlie the measurement. The first assumption is that the trapping (or de-trapping) process is completed within the period of the experiment, which in our case was 90 minutes after the gate voltage was varied. The second assumption is that barrier trapping effects can be neglected, namely that surface traps dominate the experiment. The first assumption may lead to an underestimation of the density of surface states, and the second assumption to an overestimation. We address both assumptions below.

The HFET layers, provided by NTT, were grown by metal organic chemical vapor deposition on p-type (111) silicon substrate. The barrier consists of 1 nm AlN, 25 nm $Al_{0.25}Ga_{0.75}N$ and 3 nm GaN layer, from bottom to top. The 2DEG channel is located at the AlN/GaN interface. The ohmic contacts were fabricated by e-beam evaporation of a Ti/Al/Ni/Au (20/120/40/50 nm) metal stack annealed at 825°C. Mesa isolation was performed by reactive ion etching (RIE). Two types of gated van der Pauw devices were prepared, one with and the other without the gate insulator; the former will be denoted as the MIS device and the latter as the Schottky gate device. Before the gate insulator deposition, the MIS device was dipped for 10 sec in a HF:H2O 1:50 solution, followed by plasma enhanced chemical vapor deposition of the 50-nm-thick $SiN_x$ gate insulator layer. The non-alloyed Ni/Au (30/120 nm) circular gate electrode was deposited by e-beam evaporation. Finally, polyamide was used to passivate the devices. Since excessive gate leakage introduces an error in the gated van der Pauw measurement, we have reduced the leakage compared to [17], [18], by reducing the device area and annealing the insulator layer for 10 minutes at 400°C in $N_2$ ambient conditions. The threshold voltages of the MIS and Schottky devices were -12V and -3.5V respectively.

Due to the low gate leakage, we were able to obtain the density of interface traps ($D_{it}$) across the entire bandgap, as shown in Fig. 1. The raw data, namely conductivity variation with time, is provided in Fig. 2. In Fig. 3 we compare conductivity transients in MIS and Schottky devices. In Schottky gate devices, trapping effects are due to barrier traps only, because the metal Fermi level determines the surface states occupation. The metal Fermi level is given once the energy barrier between the metal and GaN is set, and does not depend upon the applied Schottky gate voltage. Hence, the surface charge remains constant when the Schottky gate voltage is varied. The similarity between the amplitudes of the conductivity transients in MIS and Schottky devices thus indicates that barrier traps play an essential role in the gated van der Pauw experiment in MIS devices. We therefore conclude that the surface trap concentration may be lower than shown in Fig. 1.

As mentioned above, it may be possible that the period of our gated van der Pauw experiment is too short, and some electrons remain trapped at the surface following the gate voltage step, and are not monitored by experiment. To evaluate this effect, we carried out UV assisted gated van der Pauw experiments, which are similar in principle to UV assisted CV experiments [10], [13], [14]. To enable the penetration of radiation, devices with 1-nm-thick UV-transparent gate metals were fabricated, Ni/Au (0.5/0.5 nm). The gate current of these devices in the dark, and under UV illumination, is shown in Fig. 4. Both the MIS and Schottky devices exhibited significant photocurrent. The lower (but non-zero) photocurrent seen in the MIS devices is probably due to the negative energy barrier for holes between the GaN valence band and SiN, and the low hole mobility in SiN [20].

The 4.5 eV UV radiation generates a large concentration of holes that accumulate at the GaN/insulator and barrier layer in the valence band in traps [10]. To preserve charge neutrality, the UV generated charge is compensated by a variation of the 2DEG concentration, which we detect. The variation of the 2DEG concentration after 20-second-long UV illumination is presented in Fig. 5, for Schottky and MIS devices. Indeed, substantial changes are induced in the 2DEG concentration due to the net positive charge generated by the UV light. Full decay of the positive charge may take several hours. We have converted the measured 2DEG conductivity to 2DEG concentration, assuming constant mobility of $1700 \, cm^2/V \cdot s$. The justification for this assumption is deduced from the linear dependence of conductivity on gate voltage during a fast sweep, as shown in Fig. 6.

The concentration of interface states in an MIS device is evaluated from the UV assisted gated van der Pauw experiment as follows. When a negative gate voltage $V_G$ is applied before UV illumination, the 2DEG concentration varies by an amount, $\Delta n$ given by

$$\Delta n = \frac{C_{HF} V_G}{q} \quad (1)$$

(neglecting the small changes due to trapping effects) where $C_{HF}$ is the high-frequency capacitance of the gate, and $q$ is the electron charge. As a result, and as can be seen in Fig.7, the position of the quasi-Fermi level at the surface shifts from its equilibrium position by an amount $\Delta E_F$ given by

$$\Delta E_F = q^2 \Delta n \left( \frac{t_{AlGaN}}{\varepsilon_{AlGaN}} + \frac{t_{GaN}}{\varepsilon_{GaN}} \right) \quad (2)$$

where $t_{AlGaN}, t_{GaN}, \varepsilon_{AlGaN}, \varepsilon_{GaN}$ are the AlGaN barrier and GaN cap thicknesses and dielectric constants.

When the UV illumination is turned on, electrons trapped at energy levels above the quasi-Fermi level recombine quickly with the photo-generated holes [10]. We denote this electron concentration as $\delta n(V_G)$. We thus have:

$$\delta n(V_G) = \Delta E_F \langle D_{it} \rangle \quad (3)$$

where $\langle D_{it} \rangle$ is the average interface trap (or surface-donors) density across an energy range $\Delta E_F$, below the equilibrium Fermi level. The $\delta n(V_G)$ value provides the desired information on the average density of interface traps. Combining (2) and (3), we obtain that

$$\langle D_{it} \rangle = \frac{\delta n(V_G)}{q^2 \Delta n} \left( \frac{t_{AlGaN}}{\varepsilon_{AlGaN}} + \frac{t_{GaN}}{\varepsilon_{GaN}} \right)^{-1} \quad (4)$$

Ideally, one should wait after turning the UV light off for a full decay of the UV generated charge variations, except $\delta n(V_G)$, to obtain the de-trapped electron concentration. Fig. 5 indicates, however, that the decay time of the trapped positive charge generated by the UV radiation is very long. Experimental instabilities preclude waiting for such a long time. Instead, we continue by assuming that the decay rate of the UV generated charge is independent of the gate voltage. Hence, by subtracting the variation of the 2DEG concentration at zero gate voltage form the variation of the 2DEG concentration at negative gate voltages, one can obtain $\delta n(V_G)$. We denote by $\Delta n_{UV}(V_G, t)$ and $\Delta n_{UV}(V_G = 0, t)$ the time-dependent 2DEG concentration variations after UV radiation at negative and zero gate voltage respectively. Hence,

$$\delta n(V_G) \approx \Delta n_{UV}(V_G, t) - \Delta n_{UV}(V_G = 0, t) \quad (5)$$

The obtained results of the above procedure are shown in Fig.8. For the Schottky device, $\delta n(V_G = -1.5V)$ is too small to be measured, indicating that either the barrier trap concentration is low, or that the holes generated by the UV radiation drift towards the metal gate, and do not recombine with the trapped barrier electrons. In the MIS device, on the other hand, $\delta n(V_G = -7.5V)$ is small but measurable. Note that $\delta n(V_G)$ is time independent, validating the assumption underlying Eq.5. The same procedure was repeated for different gate voltages, and the results are shown in Fig. 9.

From the obtained $\delta n$ values given in Table I, the measured high-frequency capacitance $C_{HF} = 1.5 \cdot 10^{-7} F \cdot cm^{-2}$, and using the above device parameters, we obtained the average density of interface traps, presented in Table I. The values are lower by a factor of about five from the results shown in Fig. 1. The discrepancy can be explained by our findings, mentioned above, that the $D_{it}$ values shown in Fig. 1 may overestimate the real surface trap concentration.

### III. DISCUSSION

Our experimental results, in agreement with previous reports, demonstrate that the concentration of surface states in GaN is lower than required to explain the variation of the 2DEG concentration with barrier thickness by the surface donor model, and to compensate the polarization charge in III-nitrides. Thus, to account for the formation of the 2DEG and the compensation of the polarization charge, an additional surface charge should be identified. The additional charge must be "fixed"; namely, its concentration should not respond to an external excitation.

A conventional model for the additional fixed charge would be that the charge resides in an insulating layer on the surface, as suggested by [3]. For surfaces with no intentionally deposited insulator, the native oxide may include charged traps. The following arguments demonstrate, in our opinion, that this model is unlikely. Border traps in the native oxide behave similarly to surface traps, but have longer trapping and release times. They should thus be measurable, provided temperature is high enough, or UV radiation is applied. In addition, if a compensating charge of the order of $8 \times 10^{14} e/cm^2$ [2] resides in border traps in the oxide layer, any minute variation in the border trap concentration induced by surface treatments should induce a huge change in the 2DEG concentration. However, 2DEG concentration is very robust; it is not significantly affected by surface treatments. We finally note that the compensating charge is induced by polarization, which is a bulk property. Since bulk determines the compensating surface charge concentration, charge transport between bulk and border traps should take place. Hence, border trap compensating charge is not truly "fixed" and should be accessible experimentally

As charge residing in the native oxide is unlikely to supply the missing fixed charge, we suggest that an intrinsic fixed charge, unrelated to defects, should be identified. This intrinsic charge should be unaffected by surface treatments, to account for the robustness of the 2DEG concentration. To the best of our knowledge, an intrinsic surface charge was not yet identified by ab-initio calculations. In fact, it was claimed that "any finite perturbation applied to an insulating surface of an insulating crystal can have no effect at all on the areal surface charge density" [21]. Some comments about the physical nature of the suggested intrinsic polarization self-compensation (PSC) charge follow.

First, the surface of GaN is insulating. As schematically depicted in Fig. 10, at an insulating surface the position of the Fermi energy falls into an energy gap common to the bulk and surface, and the valence band is fully occupied [21]. Hence, if the surface is insulating and charged (beyond the density of charged surface traps), the integrated density of states, fully occupied by electrons, in the valence bands at the surface, does not compensate surface ion charge (namely crystal ions, i.e. gallium and nitrogen, as well as any surface oxides etc.).

Second, in thick insulating pyroelectric layers, the polarization charge is almost fully compensated at both opposing insulating surfaces. Hence, the intrinsic PSC charge approach implies that on one surface the integrated density of valence band states is lower than required to compensate the ion charge, and on the opposite surface, higher.

Third, according to older and recent calculations, the 2DEG concentration at the interface of GaN heterostructures approximately compensates the interface net polarization charge [2], [22]. Hence, no intrinsic PSC charge seems to be present at the interface between GaN and the barrier crystal. The following question should thus be considered: what is the difference between the surface of GaN, where PSC charge seems to be present, and the interface between GaN and a barrier layer, where no PSC charge exists? A plausible answer is that the position of surface ions is less rigid, compared to the position of interface ions. The sought-after intrinsic PSC surface charge in nitrides may thus be attributed to modification of the surface ion positions.

Fourth, from the above considerations, it is evident that the PSC charge fully compensates the polarization charge of thick GaN layers. On the other hand, when a sufficiently thick barrier layer is grown epitaxially on GaN, the PSC charge at the surface of the GaN layer, which becomes an interface between GaN and the grown barrier layer, is reduced to zero (see third point above). It is thus reasonable that the PSC charge concentration decreases gradually to zero with barrier thickness, accounting for the experimental variation of the 2DEG concentration with barrier thickness. The underlying physics may be the increasing rigidity of interface ion positions, as the barrier layer thickness is increased.

Finally, it is asked frequently what the origin or source of the 2DEG is. The standard answer is that surface donors are the source. Yet, we have shown here that the concentration of surface donors is much lower than the 2DEG concentration so that this answer cannot apply. We hence conclude that because surface donor concentration is too low, only valence electrons can be the source of the 2DEG.

To illustrate the above, we consider an insulating polar heterojunction between materials A and B having polarization $P_A$ and $P_B$ respectively, with neither bulk nor surface donors. We now assume that intrinsic PSC charge fully compensates the polarization charge at the surfaces, but not at the interface. At the surface of material A, the PSC charge is $-P_A$, and at the surface of material B, the PSC charge is $P_B$. The charge imbalance in the structure is thus $P_B-P_A$. Two different scenarios may take place. If $P_B-P_A$ is positive, the charge imbalance indicates that the integrated density of states across all valence bands in the entire structure is smaller than the total ionic positive charge. Excess valence electrons will then be transferred to the conduction band, and form the 2DEG at the heterojunction, with concentration $P_B-P_A$, and compensate the interface net polarization charge. On the other hand, if $P_B-P_A$ is negative, the charge imbalance indicates that the integrated density of states across all valence bands in the entire structure is larger than the total ionic positive charge. Charge neutrality thus requires that the valence band is not fully occupied, and holes in the valence band will form a two-dimensional hole gas at the heterojunction, with concentration $-(P_B-P_A)$.

## IV. CONCLUSIONS

We have shown that, in accordance with most previous reports, the surface-trap density in GaN heterostructures is much lower than required to compensate the charge of the two-dimensional electron gas. It is also much smaller than required to compensate the surface-polarization-charge. Hence, variable charge located in midgap surface-energy-states is ruled out as an explanation for the formation of the 2DEG, and the compensation of the polarization-charge at the surface of thick GaN. We claim that an extrinsic (surface oxide defect-related) charging mechanism is unlikely, and suggest that an intrinsic polarization self-compensation effect may take place. Some comments about the nature of the intrinsic polarization self-compensating surface charge are provided. We finally emphasize that our experimental results point out very clearly that fixed surface charge rather than surface states should be included in device modeling and simulation.

## V. ACKNOWLEDGMENTS

We wish to thank C. G. van de Walle, J. Kuzmik, M. Tapajna, L. Kronik, and L. Kornblum, for the fruitful discussions.

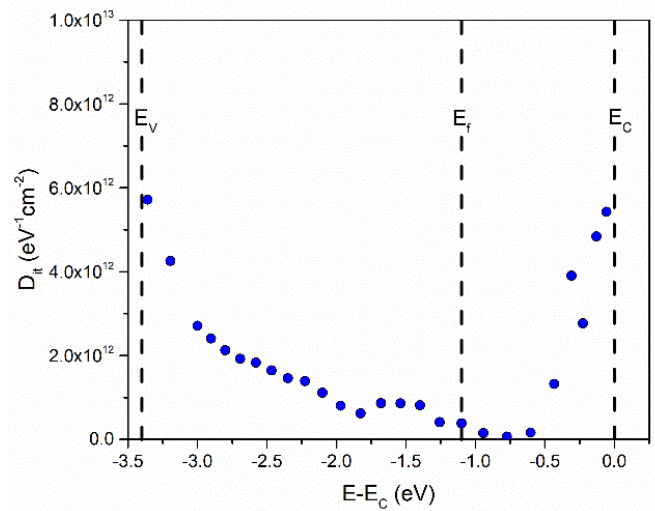

Fig. 1. The density of surface traps obtained from the gated van der Pauw experiment. The values may be an overestimation because barrier traps contribute to the conductivity transients after application of a gate voltage step.

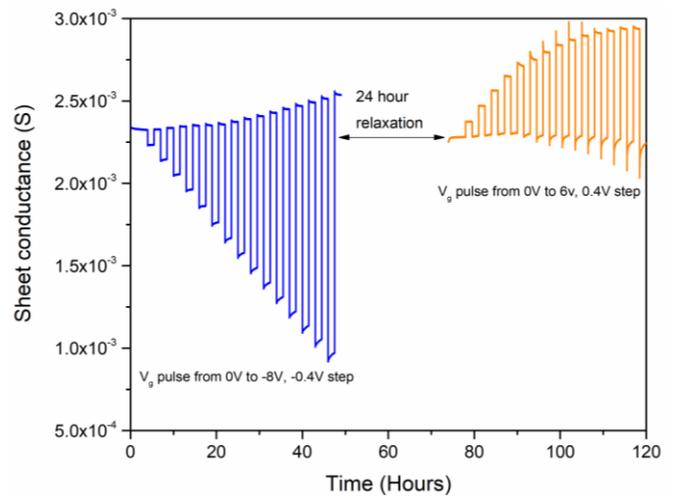

Fig. 2. Measured 2DEG conductivity versus time. The gradual variation of the conductance between gate voltage pulses (at zero gate voltage) is due to slow positive charge trapping in the insulator layer, which is compensated by an additional 2DEG charge. The density of surface states is obtained from the amplitude of the conductivity transient after a gate voltage is applied.

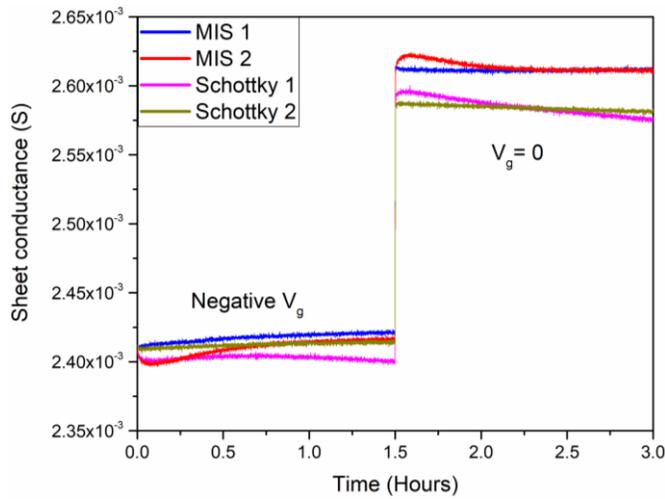

Fig. 3. Transient conductivity of two Schottky and two MIS devices. The similar variation of conductivity in the Schottky and MIS devices indicates that barrier traps play a significant role in MIS structures. The Schottky gate negative voltage is -0.4V, and the MIS gate negative voltage is -0.8V.

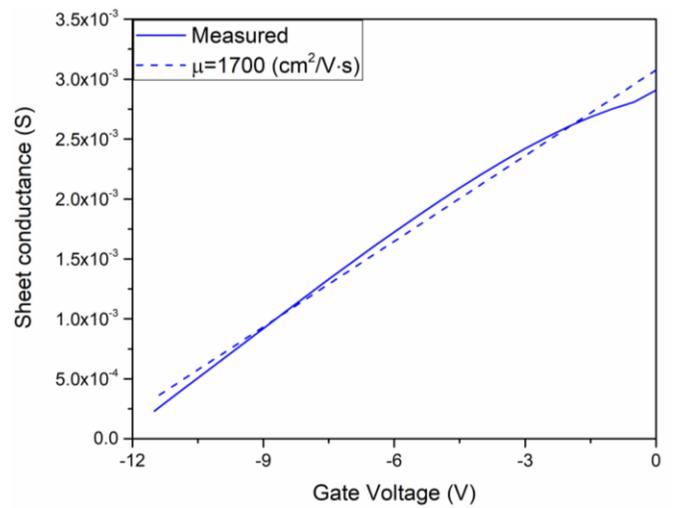

Fig. 6. Measured sheet conductance versus gate voltage in an MIS structure, and comparison to a calculation based upon the measured high-frequency capacitance and constant mobility.

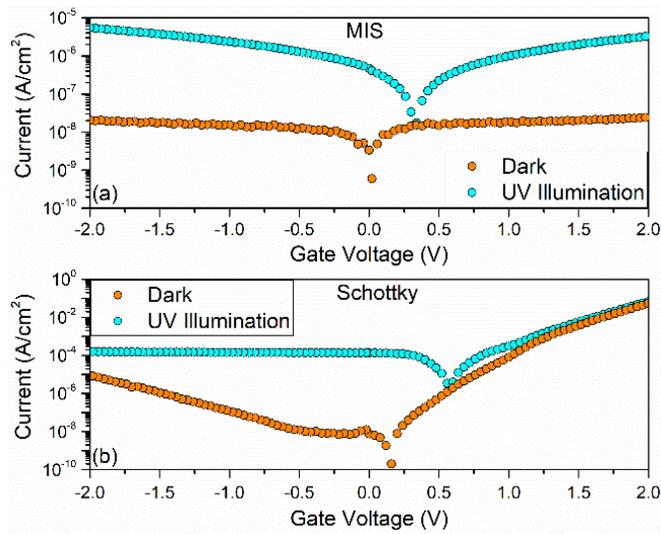

Fig. 4. UV Photoresponse of the (a) MIS gated van der Pauw structure, and (b) the Schottky gate structure.

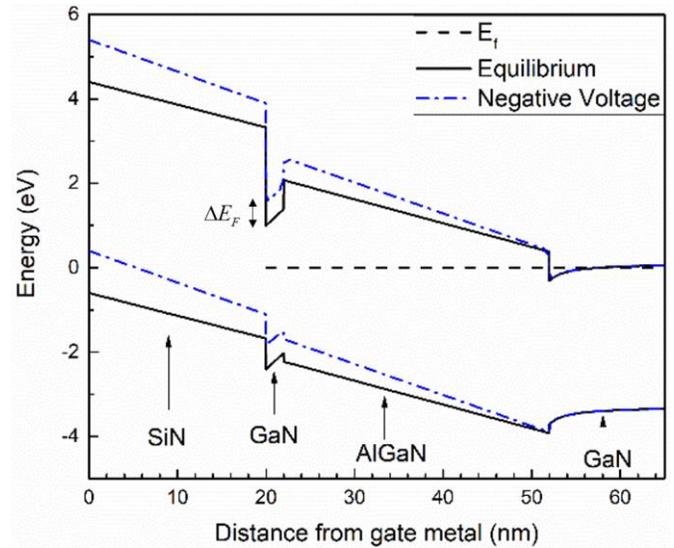

Fig. 7. Band diagram of the device at equilibrium and at a negative gate voltage.

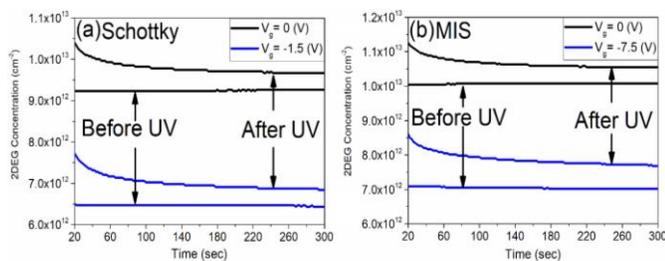

Fig. 5. Variation of the 2DEG concentration after application of UV illumination at zero and negative gate voltages for (a) a Schottky device, and (b) an MIS device.

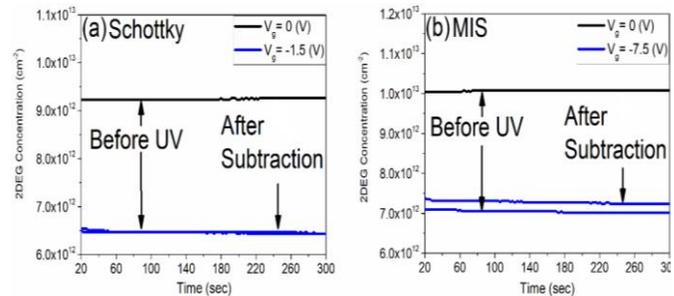

Fig. 8. Concentration variation of the 2DEG following the application of UV illumination at a negative gate voltage, after subtraction of the zero gate voltage transient for (a) a Schottky device and (b) an MIS device.

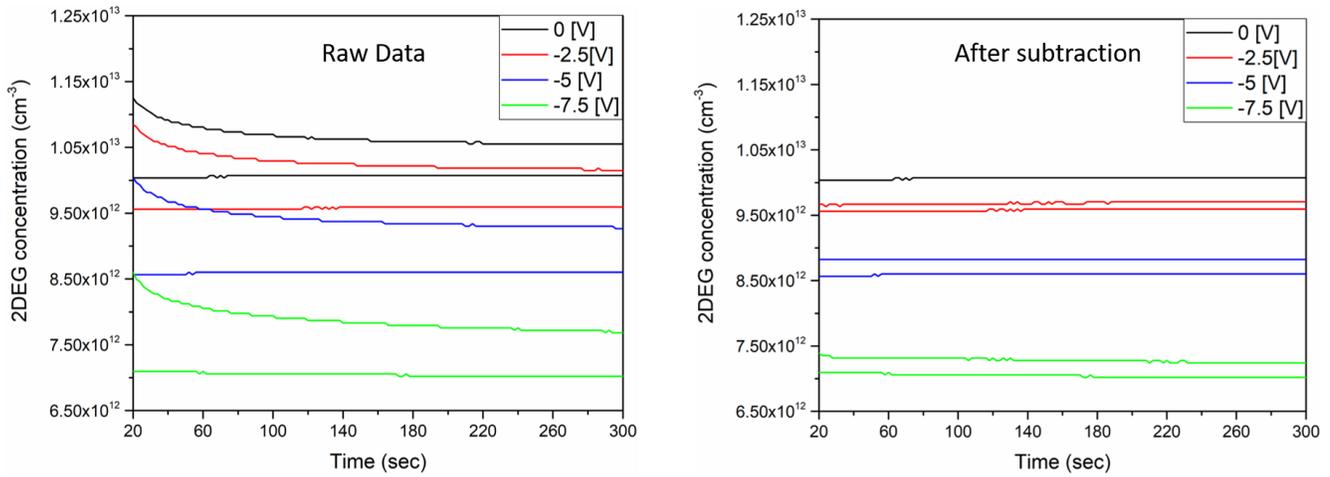

Fig. 9. UV assisted gated van der Pauw measurement results for multiple gate voltages, before and after the subtraction of the zero gate bias transient.

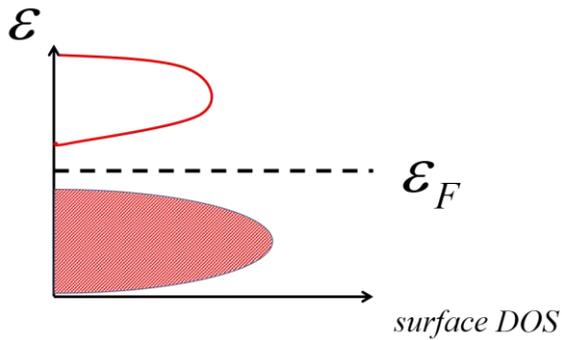

Fig. 10. The density of states at the surface of an insulating material assuming negligible concentration of localized states in the bandgap. The surface is charged when the integrated density of states below the Fermi level differs from the ion charge at the surface (namely gallium nitrogen and any oxides etc. ions).

TABLE I
UV ASSISTED GATED VAN DER PAUW RESULTS

| $V_G\ (V)$ | $\delta n\ (cm^{-2})$ | $\Delta E_F\ (eV^{-1})$ | $\langle D_{it} \rangle\ (cm^{-2}eV^{-1})$ |
|---|---|---|---|
| -2.5 | $1.06 \cdot 10^{11}$ | -0.5 | $2.12 \cdot 10^{11}$ |
| -5 | $2.27 \cdot 10^{11}$ | -1 | $2.27 \cdot 10^{11}$ |
| -7.5 | $2.34 \cdot 10^{11}$ | -1.5 | $1.56 \cdot 10^{11}$ |